# Can we use *time-resolved* measurements to get *Steady-State* Transport data for Halide perovskites?

Igal Levine, Satyajit Gupta, Achintya Bera, Davide Ceratti, Gary Hodes*, David Cahen*
*Dept. of Materials & Interfaces, Weizmann Institute of Science, Rehovot, Israel*

Dengyang Guo, Tom J. Savenije*
*Department of chemical engineering, Technical University Delft, Delft, the Netherlands*

Jorge Ávila, Henk J. Bolink*
*Instituto de Ciencia Molecular, Universidad de Valencia, Valencia), Spain*

Oded Millo, Doron Azulay, Isaac Balberg*
*Racah Institute of Physics, Hebrew University of Jerusalem, Jerusalem, Israel*

*authors for correspondence

**Abstract**

Time-resolved, pulsed excitation methods are widely used to deduce optoelectronic properties of semiconductors, including now also Halide Perovskites (HaPs), especially transport properties. However, as yet no evaluation of their amenability and justification for the use of the results for the above-noted purposes has been reported. To check if we can learn from pulsed measurement results about steady-state phototransport properties, we show here that, although pulsed measurements can be useful to extract information on the recombination kinetics of HaPs, great care should be taken. One issue is that no changes in the material are induced during or as a result of the excitation, and another one concerns in how far pulsed excitation-derived data can be used to find relevant steady-state parameters. To answer the latter question, we revisited pulsed excitation, and propose a novel way to compare between pulsed and steady state measurements at different excitation intensities. We performed steady-state photoconductivity and ambipolar diffusion length measurements, as well as pulsed TR-MC and TR-PL measurements as function of excitation intensity on the same samples of different MAPbI$_3$ thin films, and find good quasi-quantitative agreement between the results, explaining them with a generalized single level recombination model that describes the basic physics of phototransport of HaP absorbers. Moreover, we find the first experimental manifestation of the boundaries between several effective recombination regimes that exist in HaPs, by analyzing their phototransport behavior as a function of excitation intensity.

# I. Introduction

Analysis of experimental results, obtained using transient excitation methods, is often considered as equivalent to those obtained under steady-state conditions. Also in the very active field of halide perovskites, HaPs, it is assumed that the same recombination (and carrier transport) processes dominate in both types of excitations[1–7]. However, it remains unclear if the materials' electronic transport properties, deduced from the transient measurements, are indeed relevant for describing PV and LED behavior under steady-state conditions. Moreover, many studies have shown that strong light pulses can induce structural, morphological and electronic changes to an HaP, in some cases reversible[8], but in others irreversible, which calls into question the validity and relevance of the conclusions drawn from non steady-state measurements[9–15].

To examine the problem, we first survey the literature for the various electronic mobilities and carrier diffusion lengths, reported for HaPs. We distinguish between two types of measurement methods:

- *Transient,* using a pulsed light source, such as time-resolved photoluminescence (TRPL)[16], Time-Resolved THz Spectroscopy[17], Time of Flight (TOF)[18] and Time-Resolved Microwave Conductivity (TRMC)[19], and

- *Steady-state*, such as Electron Beam-Induced Current (EBIC)[20], Steady-State photoconductivity and Photocarrier Grating (SSPG)[21,22], Scanning Photocurrent Microscopy (SPCM)[23], Space Charge-Limited Current (SCLC)[24] and (DC)Hall effect measurements[6].

The most common, useful parameter that can be measured by, or extracted from one or more of the above-mentioned methods is the diffusion length of one or both of the electronic charge carrier types, electrons and holes. The principal parameter, extracted from transient measurements such as TRPL, is the radiative lifetime of the photoexcited carriers, $\tau_{rad}$, which is assumed to be the shortest one, i.e., the lifetime of the minority carriers[16]. For HaPs a wide variety of $\tau_{rad}$ values can be found in the literature, ranging from several ns to a few µs. Moreover, although ideally $\tau_{rad}$ should be measured under 1-sun equivalent excitation, as we will show later, such pulsed measurement is nearly impossible experimentally, and therefore usually $\tau_{rad}$ is measured under different pulse energies (much higher than 1-sun equivalent intensities) and is often pulse intensity- and material-dependent, making it hard to compare results from different laboratories.

Although combining the two methodologies (transient and steady-state) is not trivial, due to the very different nature of the excitation involved, it has become common practice to use a combination of both type of methods to extract the experimentally unknown charge transport parameter (diffusion length, L, or mobility, $\mu$) for the HaPs. L and $\mu$ can be related via the 1-D diffusion equation and the Einstein relation as:

$$(1)\ L = \sqrt{\frac{kT}{e}\mu\tau}$$

where $k$ is Boltzmann constant, T is the temperature and $e$ is the elementary charge. Figures 1 and 2 illustrate the differences in $L$ values between a *direct* measurement, to a *derived L* using Eq.1 and a combination of two different methods (steady-state for the mobility + pulsed for the lifetime, or even pulsed measurements for both properties). They provide a summary of reported charge carrier diffusion lengths and mobilities (minority and/or majority carriers) for MAPbI$_3$ (MAPI) and MAPbBr$_3$, respectively (for the complete list of references for these data, see the SI).

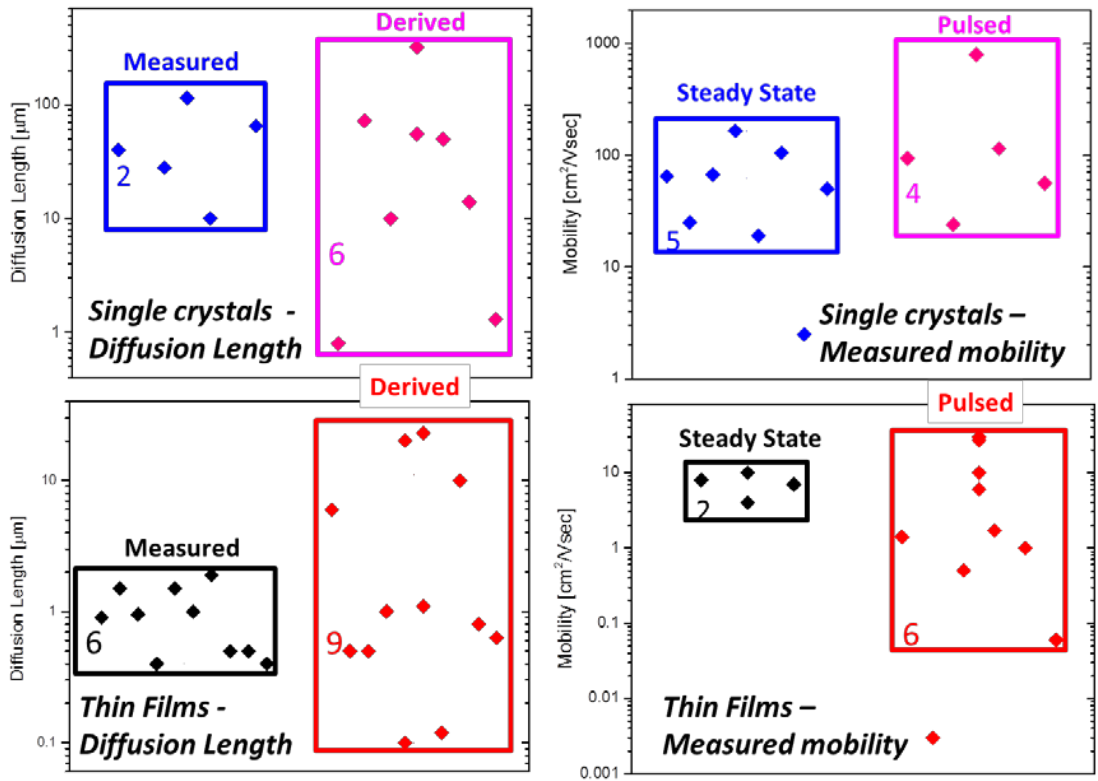

*Figure 1: Summary of MAPI transport and phototransport parameters. Each point corresponds to a reported value. The cited numbers are the total number of samples on which the results shown were derived; the outlier on the*

*steady-state single crystals mobility plot (top right) corresponds to the value obtained with the SCLC method (Shi et al.[25]). The outlier on the thin films pulsed mobility plot (bottom right) was obtained using the PhotoCELIV method (Namyoung et al.[26]) LEFT: charge carrier diffusion lengths that were directly measured, and those that were derived using the Einstein relation, for single crystals (top, left) and thin films (bottom, left). RIGHT: Mobilities reported for single crystals (top, right) and thin films (bottom right).*

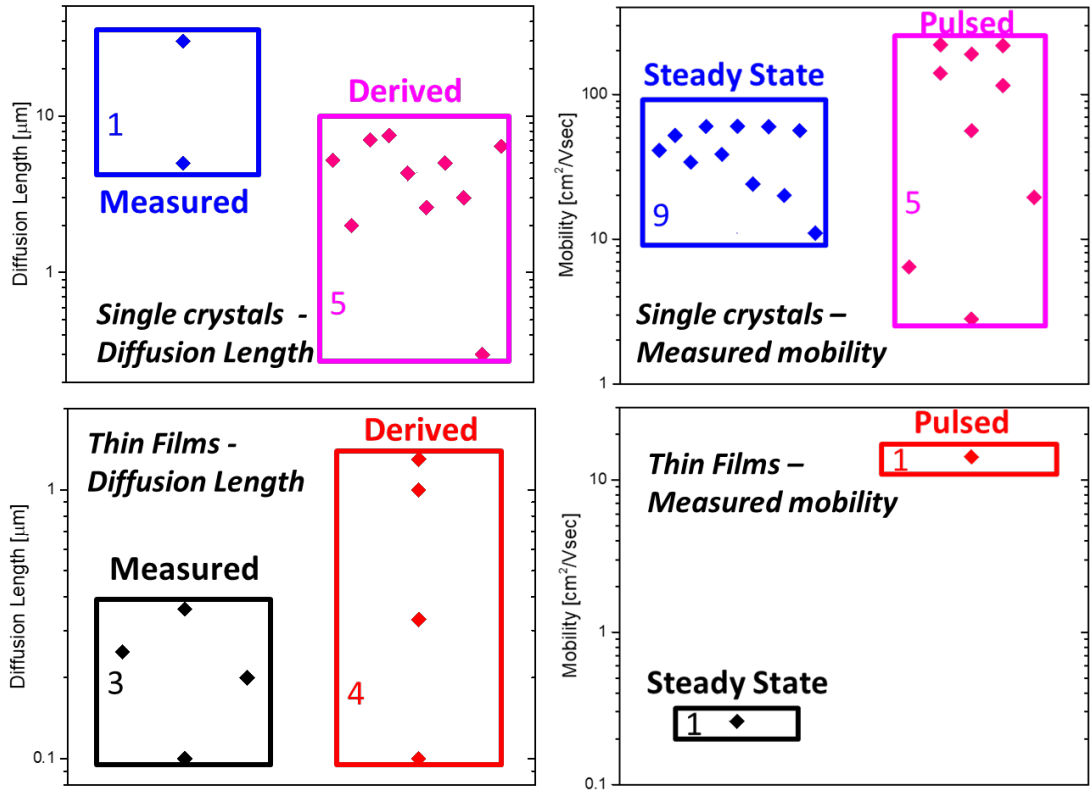

**Figure 2**: Summary of MAPbBr$_3$ transport parameters. Each point corresponds to a reported value. The cited numbers are the total number of samples on which the results shown were derived. LEFT: Directly measured charge carrier diffusion lengths, compared to those, derived using the Einstein relation, for single crystals (top, left) and thin films (bottom, left); RIGHT: Mobilities reported for single crystals (top, right) and thin films (bottom, right).

While for MAPbBr$_3$ there are currently insufficient published data regarding directly-measured single crystal diffusion length and thin film mobility (steady-state or pulsed) to draw conclusions, it can be seen that for MAPI, the spread of values for the derived carrier diffusion lengths (1-100 µm for single crystals and 0.1-20 µm for thin films) is always larger (by 1-3 orders of magnitude) than the spread of directly measured values (10-100 µm for single crystals and 0.3-3 µm for thin films). Similarly, for MAPbBr$_3$ thin films, it can be seen that the spread of values of the derived diffusion lengths (0.1-1.5 µm) is larger than the measured one (0.1-0.4 µm). The rel-atively good agreement between the *measured* L values on different samples in different laboratories for MAPI and MAPbBr$_3$, already hints that steady-state methods which measure L, might reflect more closely the actual carrier diffusion length than the wide variety of *derived* L values found in the literature.

A trivial reason for the large spread in values could be of course the quality of the measured MAPI samples, which might differ between laboratories. However, while there surely is a variation between different samples in different laboratories, we rule out this reason on the ground that if that were indeed the main cause for the observed discrepancies, a similar spread (2-3 orders of magnitude) should be seen also in the measured, steady-state values, which is not the case. To further rule out that reason, we report below on results obtained with one set of samples for different types of measurements.

Another important parameter that is often extracted from transient measurements such as THz conductivity and time-of-flight, TOF, (electrical) measurements, is the electronic carrier mobility, μ. Here, again, a wide spread of values exists in the transient measurement literature data, ranging from $10^{-3}$ to 30 cm$^2$/Vs for MAPI thin films, and 2-200 [cm$^2$/Vsec] for MAPbBr$_3$ single crystals, compared to only 1 order of magnitude variation using steady-state methods for MAPI thin films [2-10 cm$^2$/Vsec] and MAPbBr$_3$ single crystals [10-100 cm$^2$/Vsec], as shown in Figure 1. Often these two parameters, μ and τ, *deduced from transient measurements*, are combined using the Einstein relation (Eq. 1) to calculate the *steady-state diffusion lengths* of the charge carriers in the studied HaP, without justifying this transition from the pulsed to the steady-state time regime. Even if the mobility-lifetime product values are obtained using steady-state methods (such as EBIC, or steady-state photocarrier grating, SSPG) it is tempting to extract the steady-state mobility or lifetime from the experimentally derived mobility-lifetime product, using the lifetime, which can be measured directly only by a pulsed / transient method. However, as we will demonstrate here, one would need to measure *independently* and under the *same excitation conditions* as those used in the steady-state method, the mobility or the lifetime, to properly extract the other parameter using Eq. 1. To tackle this problem in a systematic manner we use two methodologies –

(1) measure different samples from different laboratories with the same methods; and

(2) measure the same samples using different (pulsed and steady-state) methods.

By performing light intensity-dependent measurements we find that the supposedly contradicting results on the carrier recombination mechanism in MAPI can be reconciled by adopting an effective single level recombination center model, and more particularly, that under the right experimental conditions, results from pulsed TRMC measurements can agree with steady-state measurements.

## II. Experimental and theoretical background

To try and understand the reasons leading to the large spreads of values, shown in Figures 1 and 2, we choose to examine these results by adopting the following approach: we compare results from pulsed and steady-state experiments by expressing the incident photon irradiation in transient experiments in terms of the equivalent carrier generation rate, $G_{eq}$, as if the pulsed incident photon flux would impinge on the sample continuously, as in steady-state, i.e., in units of [$1/cm^3$ sec]. To do so, the incident photon flux (usually expressed in photons/$cm^2$) is divided by the sample` thickness (thus assuming a uniform spatial absorption profile, which is often justified for sub-µm HaP films) and further divided by the lifetime of the charge carrier, electron or hole, as measured in the specific experiment (assuming that the pulse duration is much smaller than the probed lifetime, a condition that is met in all pulsed lifetime measurements).

In experiments where pulsed excitation is used, the measured carrier decay kinetics can vary from 1-100s of ps in THz measurements[4,7] to several or tens of µs in TRMC measurements[1,19]. These large differences in the charge carrier dynamics can be due to the very different *pulse durations* used for the excitation, ranging from fractions of ps in THz measurements[4,7] to several ns in TRMC measurements.[1,19] (An alternative, quick initial way to calculate $G_{eq}$, prior to conducting the actual experiment, is to divide the incident photon flux by the pulse duration, used for each pulsed method, rather than the measured carrier lifetime; this will result in ~ 2-3 orders of magnitudes higher $G_{eq}$ values, but we find that the general trends remain similar). While each laboratory, specialized in pulsed pump-probe measurements, has its own pump-probe setup with specific power outputs and pulse durations, the qualitative order-of-magnitude differences between the methods, applied to various HaP samples, prepared in different laboratories, can be deduced by using a range of typical laser intensities and measured lifetimes. Figure 3 shows such a comparison for 3 different types of transient methods: THz measurements (blue), TRPL (green) and TRMC (orange), illustrating the resulting equivalent steady-state generation rate, $G_{eq}$. All these ranges are compared to 1 sun for MAPI/MAPbBr$_3$ (~$10^{21}$-$10^{22}$ [photons/cm$^3$sec], shown by the horizontal black line).

One of the major drawbacks of the pulsed methods is that usually strong light pulses are required to obtain good signal-to-noise ratios[4,9]. As a result, as can be seen from Figure 3, $G_{eq}$ can change up to 3-4 orders magnitudes within the same method, depending on the experimental parameters, and reach values up to 7 (!) orders of magnitude larger than 1 sun steady-state equivalent, depending on the transient method. Thus, it is quite likely that different laboratories, using the same measurement technique, will perform measurements under different excitation conditions. Hence, this can result in large differences in the obtained values (usually of carrier lifetime), as seen in Figures 1 and 2. As a result, the quantitative, extracted values for physical parameters from such pulsed measurements can be a result of high injection, which will clearly alter the recombination kinetics of the charge carriers under study (favoring for example band-to-band bi-molecular recombination). Furthermore, due to the resulting very high values of $G_{eq}$ in the case of THz measurements, compared to 1 sun, 3$^{rd}$ order recombination pathways such as Auger recombination may become important, as was experimentally observed by Milot et al[27]. Obviously, this need not be the recombination scenario that is relevant (dominates) under steady-state, 1 sun equivalent, illumination conditions, as discussed further in the next section. In addition, or alternatively, (different) chemical changes can be induced. However, for TRMC, the resulting $G_{eq}$ can be in the regime of 1 sun – equivalent intensities (cf. Figure 3), and when studying different types of HaP samples, we expect to see a better agreement of TRMC results with those of steady-state measurements (*vide infra*) compared to those of PL or THz measurements.

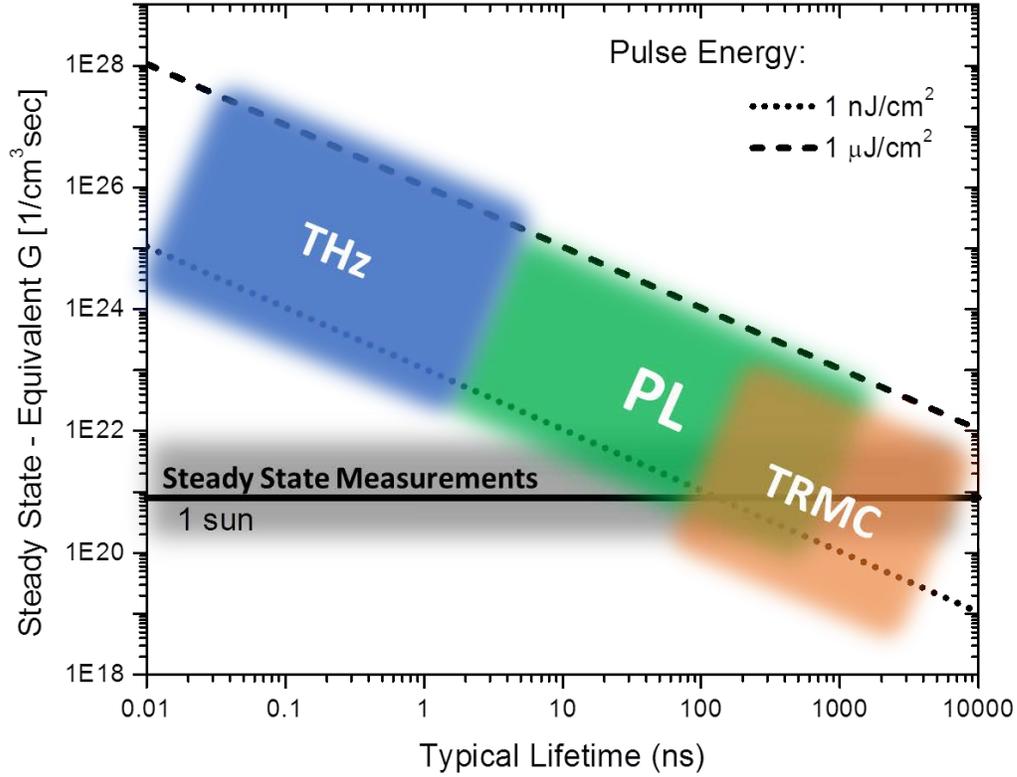

*Figure 3:* Steady-state equivalent generation rate, $G_{eq}$, as a function of the typical measured lifetime of the charge carriers in MAPI, depending on the experimental method, and the pulse energy. The horizontal black line corresponds to 1 sun equivalent of ~$10^{21}$ $1/cm^3 sec$, the grey shadow area corresponds to the experimentally accessible steady-state excitation range in our study for MAPI.

One of the simplest ways to describe the complex behavior of the different dominant recombination scenarios under different G regimes in steady-state uses a single recombination center model, as suggested by Bube[28]. We define p, n >> $p_0$, $n_0$     total, net hole and electron densities, in units of $cm^{-3}$ (for the calculations we consider the HaP to be p-type, where holes are the majority carriers, in accordance with what we found earlier[29]). Here, $p_0$, $n_0$, are the carrier concentrations in the dark.

$p_r$ ($n_r$)     density of trapped holes (electrons);

$n_{r0}$, $p_{r0}$     density of trapped electrons (holes) in the recombination level under equilibrium (dark) conditions, as indicated by the extra subscript "$_0$"; because the material is considered as p-type, $p_{r0}$ >> $n_{r0}$.

$N_r$                total number of recombination centers, $N_r = p_r + n_r$;

$C_p$ ($C_n$)      hole (electron) capture coefficient in the recombination centers;

$C_{bi}$               bi-molecular, band to band recombination coefficient;

G                generation rate, in units of [photons/cm$^3$ sec].

Note that the capture coefficient of the holes, $C_p$, essentially describes the physical process of de-trapping of an electron, i.e., release of the trapped electron to the VB, leaving a hole in the recombination center and likewise for $C_n$.

The model can be described by the following 3 equations:

$$p = n + (n_r - n_{r,0}) = n - n_{r,0} + (N_r - p_r) \quad \text{(II1)} \quad \text{(charge neutrality)}$$

$$\frac{p_r}{N_r} = \frac{C_p p}{C_p p + C_e n} \quad \text{(II2)} \quad \text{(detailed mass balance)}$$

$$G = nC_n p_r + C_{bi} np \quad \text{(II3)} \quad \text{(e-h recombination)}$$

We further recall that the common definition of the corresponding carrier lifetime is given by (see chapter 2 and 3 in Ref. [30]):

$$\tau_p = 1/C_p n_r = p/G \quad \text{(II4)}$$

and

$$\tau_n = 1/C_n p_r = n/G \quad \text{(II5)}$$

We numerically solve the above equations simultaneously to obtain n, p and $n_r$, using the following parameters: $C_n$ = 10$^{-6}$ [cm$^3$/sec], $C_p$ = 10$^{-8}$ [cm$^3$/sec], $C_{bi}$ = 5x10$^{-10}$ [cm$^3$/sec], $n_{r,0}$ = 10$^{12}$ [1/cm$^3$] and $N_r$ = 10$^{16}$ [1/cm$^3$] (the different capture coefficients were obtained from TRMC measurements, and are in good agreement with reported values[7]; for more details see section 3 of the SI). Because direct determination of n or p under photoexcitation is experimentally not practical, often only the mobility-lifetime products, $\mu\tau$, of the majority carriers, holes in our case[22], is measured via PC, and more rarely the $\mu\tau$ product of the minority carriers (electrons in our case) is measured by SSPG, or by EBIC. By using the definitions of the lifetimes (Eqs., II4, II5, and Eq. II1), we also calculate the diffusion lengths. For simplicity, we assume that the mobilities are equal, *$\mu_n = \mu_p$ = 10 cm$^2$/Vsec* and independent of G-independent, as was found by Chen et al.

using photo-Hall measurements[6], and plot the calculated carrier densities (p, n) and diffusion lengths ($L_p$, $L_n$) as function of the generation rate, G, in Figure 4. (We note that differences in the actual mobility values will only change the absolute values of L, so that the observed trends in Figures 4 will remain the same, and the only difference will be a shift up (lower mobility) or down (higher mobility) of the L-G curves in Figure 4b).

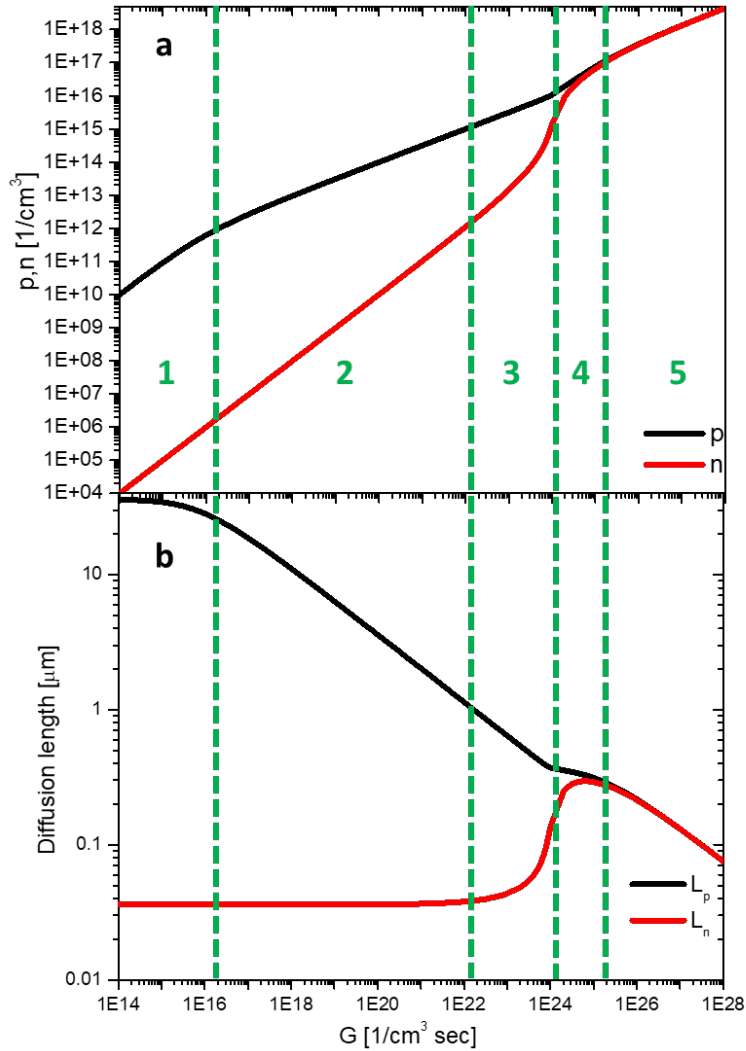

*Figure 4* : *(a) Calculated steady-state concentrations of holes (p) and electrons (n) as function of the generation rate. (b) Hole and electron diffusion lengths, $L_p$ and $L_n$, derived from the calculated hole and electron lifetimes using Eq. 1 for each generation rate by using the concentrations, given in the text (after eq. II5) and Eqs. II1 II4 and II5, at room temperature with an estimated carrier mobility of $\mu_n = \mu_p = 10\ cm^2/(Vsec)$. The green vertical lines represent transitions between different recombination regimes, as described in the main text.*

We divide the calculated results in Figure 4 into five regimes, each with a different power law dependence of the carrier concentration on the generation rate. We define these power law dependences as follows: $p \propto G^{\gamma_p}$ and $n \propto G^{\gamma_n}$. For each regime, the variation of the carrier diffusion lengths with G is different. We note that this model can be generalized to any (HaP) sample, with varying recombination center density, $N_r$, so that different capture coefficients, $C_p$ and $C_n$, can be used to describe differences between the samples; also, while the regimes discussed below will be present, the boundaries between them will shift, *depending on the $N_r$ used*. Thus, Figure 4 can explain the different experimental results in the literature for HaP transport properties as function of G, as long as the translation of G from pulsed to steady-state measurements ($G_{eq}$) is done in the manner, detailed in the introduction.

How can we distinguish qualitatively between the five regimes, from low to high G values? To that end we use the relative densities of the charge carriers, n and p, with respect to the carrier densities in the recombination level, $n_r$ and $p_r$ (for details see section S2 of the SI).

### *Regime 1:*   $G < 10^{16}$ [1/cm$^3$sec],   n, p << $n_{r0}$, $p_{r0}$

At low G values, the photogenerated hole and electron concentrations are lower than the initial (dark) concentration of the electrons and holes that occupy the recombination level and, therefore, $n_r \approx n_{r0}$, and $p_r \approx p_{r0}$, (where the subscript "$_0$" denotes the concentration of trapped carriers in the dark); thus, the lifetimes (or diffusion lengths) of both electrons and holes remain relatively constant as G changes in this regime.

### *Regime 2:*   $10^{16} < G < 10^{22}$ [1/cm$^3$sec],   n, p << $p_{r0}$, p→$n_r$

At intermediate G values, the concentration of photogenerated holes and electrons exceeds that in the dark, but still $p_r >> n_r$ and, therefore, in practice, all the photoexcited electrons get captured immediately in the recombination centers, so that $n_r$ increases with increasing G. As a result, $\tau_p$ (or $L_p$) decreases with increasing G. However, since still n << p (following Eqs. II4 & II5 and because $p_r >> n_r$), we find that $p_r \approx p_{r0}$ and $\tau_n$ (or $L_n$) remains constant with respect to G, similar to regime 1. However, due to the charge neutrality, $p \approx n_r$ (see Eq. II1) and thus we find that $p \propto G^{1/2}$ i.e., $\tau_p \propto G^{-1/2}$ and $L_p \propto G^{-1/4}$. Based on this analysis we suggest that this transition from regime 1 to 2 is the one that was observed by Chen et al.[6] for solution-grown MAPI films and by Yi et al. for Br-based HaP single crystals[31].

*Regime 3:*     $10^{22} < G < 10^{24}$ [1/cm³sec], $p \gg n_r$, $n \rightarrow p_r$, $n_r \rightarrow p_r$

In this regime, the concentration of trapped electrons, $n_r$ increases so that it is not small compared to $p_r$, but rather becomes comparable to $p_r$. Since $N_r = p_r + n_r$ is constant, $p_r$ will start decreasing as G (and $n_r$) increase, and therefore $\tau_n$ will *increase* with increasing G. We note that only in this "special" regime, the minority carrier diffusion length, $L_n$ in our case, will increase with increasing G. Based on this analysis we suggest that such a phenomenon was observed experimentally by Kedem et al.[32] for Br-based HaP solar cells, using the EBIC method.

*Regime 4:*     $10^{24} < G < 10^{25}$ [1/cm³sec], $n \approx p$

At specific G values, in our case $10^{24} < G < 10^{25}$, the concentration of the minority carriers (electrons) approaches that of the majority carriers (holes), and $n \approx p$. Under these conditions, the concentrations of $p_r$ and $n_r$ will remain constant (for the mathematical derivation, see section S2 of the SI), and the final outcome will be trap-assisted bi-molecular recombination (non-radiative).

*Regime 5:*     $G > 10^{25}$ [1/cm³sec], $n = p \gg n_r, p_r$

At high G values, in our case $G > 10^{25}$, the concentrations of photoexcited holes and electrons exceed that of the recombination centers, $N_r$; when $p$ and $n$ are large enough (see green dashed line that distinguishes between regime 4 and 5, that is for $p, n > 10^{17}$ cm$^{-3}$ in our case) the band-to-band recombination (electron from the CB recombines directly with a hole in the VB) becomes dominant over recombination via gap state levels. The interplay between the two processes when $n = p$ (radiative, band to band recombination vs. non-radiative recombination via gap state levels) will be determined by the different capture coefficients, $C_n$, $C_p$ and $C_{bi}$. Since $C_{bi}$ is relatively small for the HaP, similar to GaAs[33,34], we expect that it will be dominating for $G > 10^{25}$ [1/cm³sec]. In this regime, since the dominant recombination path is the bi-molecular radiative one, the lifetimes of the holes and electrons will be equal and will decrease as G increases (due to higher probability for recombination events as more e-h pairs are formed), and the PL intensity should increase substantially. We note here, however, that in the lower G regimes (here regimes 1-3), which are relevant for steady-state illumination conditions close to, or below 1 sun (the practical operating conditions for PV without concentration), radiative recombi-nation of electrons and holes in MAPI is a very low fraction of the total recombination events,[22,35,36] < 5% (although this fraction can be increased by certain surface treatments, as was recently reported[37]). Therefore, if any conclusions are to be drawn from TRPL for steady-state measurements in regimes 1-3, one should keep in mind that these conclusions apply to this very small fraction of electron-hole pairs, and *not* to the majority of the charge carriers undergoing recombination in these ranges of G values ($G_{eq}$ > $10^{24}$ [1/cm$^3$sec]).

*Table 1:*
*Summary of the predicted majority (p) and minority (n) carrier concentrations power law exponents and diffusion lengths` behavior, as a function of the studied excitation range, for the single level recombination center model*

|  | G regime | | | | |
|---|---|---|---|---|---|
|  | **1** | **2** | **3** | **4** | **5** |
| $\gamma_p$ | 1 | 0.5 | 0.5 | 1 | 0.5 |
| $L_p$ (or $\tau_p$) | ~constant, $L_p \gg L_n$ | ↓ | ↓ | ~constant, $L_p \approx L_n$ | ↓ |
| $\gamma_n$ | 1 | 1 | > 1 | 1 | 0.5 |
| $L_n$ (or $\tau_n$) | ~constant $L_n \ll L_p$ | ~constant | ↑ | ~constant, $L_p \approx L_n$ | ↓ |

*we note that all up/down arrows apply to both $L_p$ and $\tau_p$, and the carrier mobility is presumed to be G-independent.

Table 1 summarizes the power law exponents and changes in the carrier diffusion lengths in each regime with increasing generation rate (for the derivations of the quantitative determination of $\gamma_p$ and $\gamma_n$, see section S2 in the SI). As can be seen from the table, several combinations of $\gamma_p$ and $\gamma_n$ exponents are possible, depending on the G regime.

A very important conclusion can be drawn immediately from Table 1 - experimental determination of the majority carrier power law dependence, $\gamma_p$, *alone*, from simple photoconductivity measurements for example, i.e., $\sigma_{ph} \propto G\mu_p\tau_p \propto G^{\gamma_p}$, as is commonly done for the HaP`s[6,38,39], *does not* allow unique determination of the relevant G regime. Hence, the dominant recombination path cannot be determined, because for $\gamma_p = 0.5$ for example, regimes 2, 3 and 5 are possible, and without experimentally determining also $\gamma_n$, *in the same G range in which the $\gamma_p$ was determined* (by SSPG or EBIC under illumination for example), the relevant G regime *cannot* be determined uniquely. Regimes 1 and 4 are exceptions, as exponents $\gamma_p$ and $\gamma_n$ of both the majority and minority carrier power density on the generation rate are the same for these two regimes. However, in regime 1 most recombination centers would be empty of minority carriers (electrons in the case we consider here) and therefore $L_p \gg L_n$, while in regime 4 the carrier diffusion lengths should be roughly similar (up to the root-squared ratio of the mobilities), due to the fact that in this regime, $\tau_p \cong \tau_n$, as explained above. In reality the transitions between the different regimes are not abrupt, but continuous, leading to experimental values of $\gamma_p$ and $\gamma_n$ (except in regime 3) that can be anywhere between 0.5 to 1. We note in passing that in practice, values lower than 0.5 are found for cases that cannot be described by the simplified single level model, considered here, as in cases where two or more types of recombination centers are active in the recombination process.

## III. Experimental results and discussion

Having clarified the different recombination scenarios in our simple model, we turn now to our experimental results. We compare, in light of the trends shown in Figure 4 and Table I, our steady-state results with results that we obtained by pulse measurements on the *same* MAPI samples. In particular, we follow the observed dependences on the different G or $G_{eq}$ regimes in which each method is performed. We start by measuring the phototransport properties by means of SSPG and PC of two different types of MAPI samples (3 of each type) from two different laboratories, one being solution-processed MAPI[22] (hereinafter termed simply as "MAPI"), and the other evaporated MAPI (e-MAPI), which is known to yield high-efficiency (vacuum-deposited) MAPI solar cells[40]. As can be seen from Figure 5a, for the solution-processed MAPI at generation rates in the range of $10^{20}$-$10^{22}$ [1/cm$^3$sec], the hole diffusion length decreases from ~1.5 to ~0.4 μm with increasing G, while the electron diffusion length remains fairly constant or slightly increases around 0.3-0.4 μm. Thus, comparing the results with the predictions of Figure 4, we conclude that these phototransport measurements were performed in regime 2, probably approaching regime 3 close to G~$10^{22}$ [1/cm$^3$sec]. In contrast, the dependence on G for the steady-state phototransport-derived diffusion lengths of the holes and electrons in the e-MAPI films (Figure 5a, red) behave the same way: both decrease with increasing G. This suggests that for the e-MAPI, the phototransport measurements were carried out in regime 5. If so, according to Figure 4 and table 1, we may conclude that for the e-MAPI, $\tau_p \cong \tau_n$. The difference between the diffusion in the two cases is due to the higher mobility of the holes (note that in Figure 4, we assumed that $\mu_n = \mu_p = 10\ cm^2/Vsec$). The $L_h \approx 3L_e$ result in Figure 5a suggests that for the e-MAPI case, $\mu_p = 9\mu_n$. Support for our conjecture that in the e-MAPI samples we probe the charge carrier dynamics in regime 5, comes from our TRPL measurements. As shown in Figure 5b, the TRPL response is ~10 times larger in the e-MAPI samples than in the MAPI samples and supports the conclusion, derived from the steady-state phototransport measurements (Figure 5a), that band-to-band radiative recombination is much more dominant for the e-MAPI sample than for the MAPI sample.

Another interesting observation from the TRPL result is that the e-MAPI TRPL lifetime (~10 ns) is lower by a factor of about ~50 than that of the solution-processed MAPI film (~500 ns). In spite of the significant differences between the TRPL lifetimes of the two samples, from the phototransport measurements we found that both types of films exhibit similar µτ products ([4-11]X $10^{-7}$ $cm^2$/V), in agreement with what we already reported for the solution-processed MAPI films[22]. Thus, although in the literature there is an often implicit, common assumption that longer radiative lifetimes result in higher PV quality films and better PV conversion efficiencies [15,41–44], we do not find such direct correlation between measured radiative lifetimes and phototransport properties. In view of these results we suggest that the reason is the very different nature of the two methods, phototransport vs. TRPL. First, photoluminescence efficiencies are generally low; therefore, they are representative of only a small fraction of the excited carriers. Second, while in transient TRPL measurement only the charge carriers, which undergo radiative band-to-band recombination, are probed, in phototransport measurements, these carriers are exactly the relatively fewer ones that *do not* contribute to the phototransport signal (or to the photocurrent in an operating solar cell under illumination)[36]. Therefore, *the carrier lifetimes extracted from a single TRPL measurement are less, if at all relevant,* to the standard operating conditions of MAPI as a light absorber in a solar cell device. Moreover, if we assume that the decay time is indeed the relevant time scale for charge carriers in the bands, we will also have to assume that the mobilities of the evaporated e-MAPI films is significantly higher than those of the MAPI. We ruled out this possibility by performing TRMC measurements, as shown in Figure S1 in the SI.

The differences in the $G_{eq}$ values for the TRPL measurements shown in Figure 5b for the MAPI ($G_{eq}$ = (1.6 X $10^{15}$ 1/cm$^3$) / 500 ns = 3.3 X $10^{21}$ 1/cm$^3$sec) and the e-MAPI sample ($G_{eq}$ = (1.6 X $10^{15}$ 1/cm$^3$) / 10ns = 1.7 X $10^{23}$ 1/cm$^3$sec) demonstrate also that the charge carrier dynamics for the e-MAPI is probed at excitation intensities that are very different for the steady-state measurements, such as SSPG and PC, and for TRPL, as suggested in Figure 2. Thus, if the probed quantity (lifetime/diffusion length) is not measured under 1-sun equivalent excitation intensity and where the measured quantities are extracted from results obtained by *methods that do not depend on phototransport of the charge carriers*, (e.g., by radiative decay), then combining two of the three measured quantities (lifetime/ mobility/ diffusion length), to derive the third, unknown parameter from it, can easily result in incorrect results. In such a case the burden of proof is on those that choose to use such a procedure. This problem may explain (some of) the large spread of <u>derived L</u> values, that is shown in Figure 1.

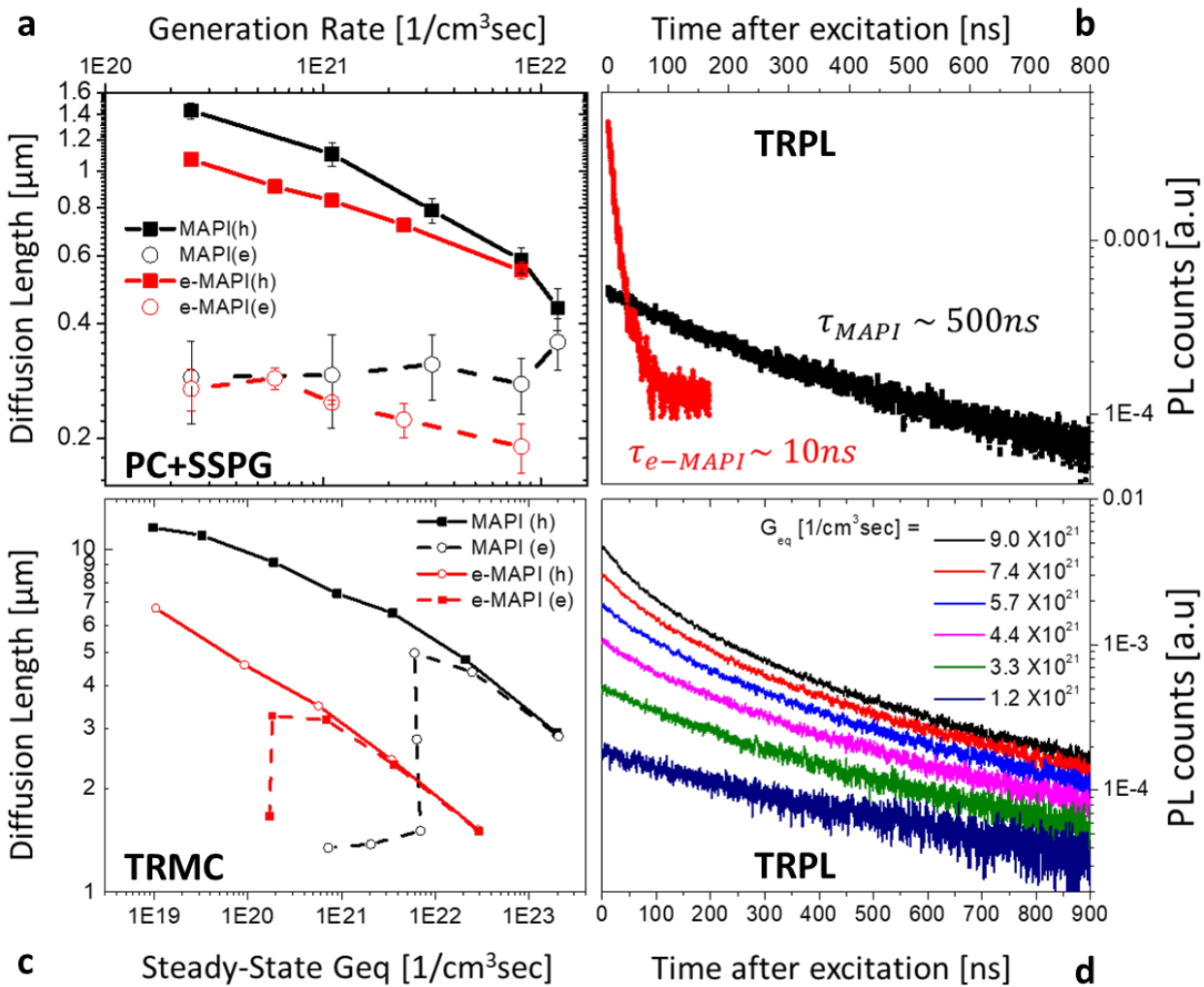

***Figure 5:*** *typical (a) holes (filled squares) and electrons (hollow circles) diffusion lengths from SSPG and PC, as function of the generation rate, G, for solution-processed MAPI (black) and e-MAPI (red); (b) TRPL results, under 640 nm excitation and 770 nm detection, for MAPI (black) and e-MAPI (red); (c) holes (filled squares) and electrons (hollow circles) diffusion lengths for solution-processed MAPI (black) and e-MAPI (red) derived from TRMC measurements, as a function of $G_{eq}$; (d) TRPL response under different pulse energies for solution-processed MAPI.*

It is now clear that the charge carrier dynamics, i.e. the lifetime of the minority carriers, probed using TRPL, is not necessarily the one that should be used to properly derive the value of L that is relevant for a device working in steady-state; thus, it need not be an indicator for the photoelectronic quality of HaP films. At the same time, according to our expectation from Figure 2, the TRMC measurements can yield results under conditions closer to those of solar cell operation and, thus, should yield results that are similar to those from steady-state methods. Furthermore, as Figure 5a suggests that for solution-processed MAPI at $G\sim10^{22}$ 1/cm$^3$sec we approach regime 3, we set out to try and measure the carrier diffusion lengths beyond $G\sim10^{22}$ 1/cm$^3$sec by TRMC to look for the transition to regimes 4 or 5, predicted in Figure 4. It is obvious that these regimes cannot be obtained by steady-state methods since the material will be structurally damaged or evaporate in the worst case[8,11] and its electronic properties will drastically change[12,13].

TRMC results for MAPI and e-MAPI are shown in Figure 5c (for details regarding the extraction of the hole and electron diffusion lengths from the raw data, see section S3 in the SI and Refs.[1,37]). Under conditions equivalent to generation rates of $10^{20}$-$10^{22}$ 1/cm$^3$sec, the trends arising from the TRMC results are in good agreement with the steady-state results (Figure 5a) for both types of samples. For both film types, the hole diffusion length decreases with increasing G, while, similar to the steady-state results, for the MAPI the electron diffusion length remains rather constant; for the e-MAPI, although the electron diffusion length decreases with light intensity. Thus, the TRMC results suggest that in this G range, we are in regime 2 for the MAPI, and regime 5 for the e-MAPI. This serves as a good demonstration to how TRMC measurements can agree well with steady-state measurements. Indeed, Semonin et al.[23] have recently shown such an agreement for MAPI single crystals by comparing TRMC and SPCM results (although in that case the agreement was for a single value, rather than a trend with light intensity as we show here). However, although the trends in the phototransport measurements and the TRMC are similar, it can be seen that in our case, the diffusion length values obtained from the TRMC measurements are roughly 5-10 times *higher* than those measured in steady-state. This discrepancy could arise from several factors such as:

(1) the exact value of the mobility of electrons and the holes is not known. As a first order approximation they were assumed to be equal and the values, used for plotting Figure 5c, were taken as half the sum of the mobilities (see Figure S1 and section S3 of the SI). This approximation is questionable, as the phototransport results suggests that for the e-MAPI, $\mu_p = 9\mu_n$;

(2) several inherent differences between the two methods. While SSPG measures phototransport on a large length scale, i.e., across grain boundaries, TRMC is a local, non-contact method, operated under open circuit (compared to the short circuit conditions in the SSPG and PC measurements), and hence larger carrier lifetimes are expected where no charge extraction occurs. Furthermore, the TRMC the signal might be dominated by charge carrier dynamics within the bulk of small single crystallites, which resemble more the transport properties of single crystals rather than those of thin polycrystalline films. Hence, in view of reports on grain boundary effects in HaPs[45–47], larger mobilities and diffusion lengths are obtained. If this interpretation is correct the TRMC measurements can reveal what is the "potential" quality of the film, e.g., if by a different film processing route less or more electronically benign grain boundaries were formed or eliminated.

More importantly, in the TRMC results of the MAPI, when approaching $G_{eq} = 10^{22}$ 1/cm$^3$sec, the electron diffusion length increases from 1 μm to ~5 μm, becoming equal to the hole diffusion length, where n=p, and they then decrease together as $G_{eq}$ approaches $10^{23}$ 1/cm$^3$sec, due to the crossover to regime 5. Thus, in the TRMC measurements, thanks to the larger range of accessible G values (4 orders of magnitudes compared to 2 in the phototransport measurements), not only regimes 2-3 are observed and agree well with the results obtained from the phototransport measurements, but all the regimes from regime 2 to regime 5, as shown in Figure 5c. are experimentally observed. (We further note here that in our analysis we neglect Auger recombination, since it is not relevant in the $G_{eq}$ range studied here, as explained earlier in the text). For the e-MAPI, the TRMC results suggest that the crossover to regime 5 is roughly at $G_{eq} = 10^{20}$ 1/cm$^3$sec, *2 orders of magnitudes lower than that observed for the MAPI*. The TRMC results are then in very good agreement with the model presented in Figure 4 in the previous section, sug-gesting that the simple model, comprising a single level recombination center and bimolecular recombination, can describe well the recombination kinetics in MAPI and e-MAPI films. To further verify that for the solution-processed MAPI sample there is crossover from regime 3 to 4 at the $G_{eq}$ range of $10^{21}$-$10^{22}$ 1/cm$^3$sec as the TRMC results suggest, we measured the TRPL response of the MAPI film for several excitation intensities within this $G_{eq}$ range. The results are shown in Figure 5d and the dependence on the TRPL decays, as a function of excitation intensity, suggests that there is a change in the decay mechanism as the intensity is increased, switching from monomolecular recombination at low intensities, to bi-molecular recombination, at higher intensities (1/t to $e^{-1/t}$ t-dependence). This suggests that the TRPL measurements probe a transition from regime 3 to regime 4 in our MAPI sample, which is in agreement with the TRMC results (we could not repeat the same exercise for the e-MAPI films, since reaching G values in the range of $10^{19}$-$10^{20}$ 1/cm$^3$sec would require excitation pulses with extremely low photon dose, yielding TRPL signals well below the sensitivity of our experimental setup).

Combining our steady-state and pulsed experimental results, we learn that the major difference between the solution-processed MAPI and the e-MAPI samples is the range of G values in which the crossover between regimes 3 to 4 occurs, where $n \approx p$ (which we term hereinafter as $G_{co}$). This transition occurs at $G_{co} \approx 10^{22}$ 1/cm$^3$sec for the solution-processed MAPI sample, but for the e-MAPI sample this transition is observed at $G \approx 10^{20}$ 1/cm$^3$sec, suggesting that for e-MAPI, the concentration of recombination centers is much smaller than in the solution-processed MAPI samples.

To illustrate the difference in $G_{co}$ for the two types of samples, we performed simulations using the same parameters as in Figure 4, but changed the concentration of the recombination centers, $N_r$ in the range of $10^{11}$-$10^{18}$ cm$^{-3}$, and extracted $G_{co}$ for each $N_r$. As expected, $G_{co}$ increases with increasing $N_r$, as shown in Figure 6.

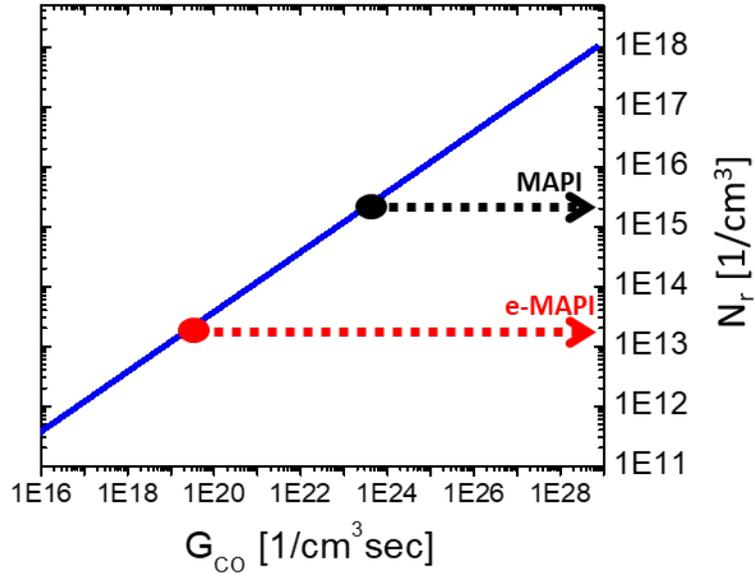

***Figure 6:*** *Calculated $G_{CO}$ values for different $N_r$ values, $G_{CO}$ corresponds to the G in which $n \cong p$. The colored circles correspond to the experimentally found $G_{CO}$ for the MAPI (black) and the e-MAPI (red).*

Therefore, we conclude that for solution-processed MAPI, $N_r \approx 10^{15}$ cm$^{-3}$ (shown in black in Figure 6), but for e-MAPI (red), $N_r \approx 10^{13}$ cm$^{-3}$. This result implies a significantly lower density of defects that serve as recombination centers in e-MAPI than in solution-processed MAPI. The question that arises next is, if indeed this is the case, why are the diffusion lengths (i.e., the $\mu\tau$ products) of both films similar, as shown in Figure 5a, because, from Figure 6 we would expect the diffusion lengths in the e-MAPI to be higher than those in the MAPI. We suggest that the reason for that the $\mu\tau$ products of both films are similar, is a combination of:

-a-    On the one hand, a lower $N_r$ should yield a larger $\mu\tau$;

-b-    On the other hand, stronger PL yield under 1-sun conditions will result in more e-h pairs undergoing radiative recombination, for the e-MAPI film (and, thus, will not contribute to phototransport, which is what is measured in the SSPG and PC) than in MAPI films. The reason is, as explained above, that band-to-band *radiative* recombination dominates in the e-MAPI, because, as shown in Figure 5a, and simulated in Figure 4, regime 5 is the relevant one for the e-MAPI films under 1 sun conditions, and in that regime >50% of the e-h pairs undergo radiative recombination, compared to < 5% for MAPI films[36].

## IV. Conclusions

While there is a large spread of reported values of mobilities and diffusion lengths, using pulsed methods, especially for the MAPI, we find that this spread is not only due to the variability in sample quality and preparation conditions, but also to the interpretation of results that were obtained by different techniques that are used in the measurements. This is in particular due to the different excitation conditions, which vary between the various pulsed methods and can differ greatly in terms of equivalent intensity (and generation rate) from the steady-state ones. The latter realization of the wide differences calls for an attempt to evaluate the information that is derived by the various techniques, which is frequently contradictory and thus not amenable for the determination of the phototransport parameters. A conspicuous example is the attempt to use information, obtained with high-power short-time excitation pulses to derive parameters relevant to steady-state operation that occurs at much lower excitation power. We therefore revisited pulsed excitations, taking into account the lifetime of the charge carriers and the generation rate equivalents of those to the steady-state excitation conditions. To do that we suggested a simple calibration for the comparison of the two types of excitations. To see then the benefits of this calibration we applied a simple model with a single recombination center, which is found to explain well the photophysical properties related to electronic charge transport in different HaP absorber materials. Using that model and mapping the pulse excitation intensity on the steady-state power scale, we compared between our experimental results, obtained by pulsed and steady-state excitation measurements on polycrystalline films, prepared in different laboratories via different preparation routes. We found a relatively good agreement between TRMC and steady-state measurements. However, we also found that depending on the studied HaP sample, $2^{nd}$ order processes, such as bi-molecular radiative recombination, may be the dominant processes contributing to the observed signal in TRPL measurements, yet they are not necessarily relevant for the standard operating conditions of the HaP as absorbers in solar cell devices. We further suggest that experimental determination of the exact dominant recombination mechanism in HaP materials should include measurements of *both* the majority and minority carrier` phototransport properties and that these should be carried out under illumination conditions that are as close as possible to 1 sun, preferably with some intensity vari-ation to allow the determination of the carrier generation regime at which the data were obtained.

## Acknowledgements

We thank Dan Oron from the WIS for fruitful discussions. GH and DC thank the SolarERAnet program HESTPV, via the Israel ministry of Infrastructure, for partial support. At the Hebrew University this work was supported in part by the Harry de Jur Chair in Applied Science (O.M.) and the Enrique Berman Chair in Solar energy research (I.B.). HB acknowledges support from the Spanish Ministry of Economy and Competitiveness (MINECO) via la Unidad de Excelencia María de Maeztu MDM-2015-0538, MAT2017-88821-R, PCIN-2015-255.

# References

[1] E.M. Hutter, G.E. Eperon, S.D. Stranks, and T.J. Savenije, J. Phys. Chem. Lett. **6**, 3082 (2015).

[2] S.D. Stranks, G.E. Eperon, G. Grancini, C. Menelaou, M.J.P. Alcocer, T. Leijtens, L.M. Herz, A. Petrozza, and H.J. Snaith, Science **342**, 341 (2013).

[3] G. Xing, N. Mathews, S. Sun, S.S. Lim, Y.M. Lam, M. Gratzel, S. Mhaisalkar, and T.C. Sum, Science (80-. ). **342**, 344 (2013).

[4] C. Wehrenfennig, G.E. Eperon, M.B. Johnston, H.J. Snaith, and L.M. Herz, Adv. Mater. **26**, 1584 (2014).

[5] E. Alarousu, A.M. El-Zohry, J. Yin, A.A. Zhumekenov, C. Yang, E. Alhabshi, I. Gereige, A. AlSaggaf, A. V. Malko, O.M. Bakr, and O.F. Mohammed, Doi.org 4386 (2017).

[6] Y. Chen, H.T. Yi, X. Wu, R. Haroldson, Y.N. Gartstein, Y.I. Rodionov, K.S. Tikhonov, A. Zakhidov, X.-Y. Zhu, and V. Podzorov, Nat. Commun. **7**, 12253 (2016).

[7] R.L. Milot, G.E. Eperon, H.J. Snaith, M.B. Johnston, and L.M. Herz, Adv. Funct. Mater. **25**, 6218 (2015).

[8] D.R. Ceratti, Y. Rakita, L. Cremonesi, R. Tenne, V. Kalchenko, M. Elbaum, D. Oron, M.A.C. Potenza, G. Hodes, and D. Cahen, Adv. Mater. **30**, 1706273 (2018).

[9] D.W. deQuilettes, W. Zhang, V.M. Burlakov, D.J. Graham, T. Leijtens, A. Osherov, V. Bulović, H.J. Snaith, D.S. Ginger, and S.D. Stranks, Nat. Commun. **7**, 11683 (2016).

[10] X. Wu, L.Z. Tan, X. Shen, T. Hu, K. Miyata, M.T. Trinh, R. Li, R. Coffee, S. Liu, D.A. Egger, I. Makasyuk, Q. Zheng, A. Fry, J.S. Robinson, M.D. Smith, B. Guzelturk, H.I. Karunadasa, X. Wang, X. Zhu, L. Kronik, A.M. Rappe, and A.M. Lindenberg, Sci. Adv. **3**, e1602388 (2017).

[11] X. Wu, H. Yu, L. Li, F. Wang, H. Xu, and N. Zhao, J. Phys. Chem. C **119**, 1253 (2015).

[12] D.J. Slotcavage, H.I. Karunadasa, and M.D. McGehee, ACS Energy Lett. **1**, 1199 (2016).

[13] E. Mosconi, D. Meggiolaro, H.J. Snaith, S.D. Stranks, and F. De Angelis, Energy Environ. Sci. **9**, 3180 (2016).

[14] E.T. Hoke, D.J. Slotcavage, E.R. Dohner, A.R. Bowring, H.I. Karunadasa, and M.D. McGehee, Chem. Sci. **6**, 613 (2014).

[15] T.A. Berhe, J.-H. Cheng, W.-N. Su, C.-J. Pan, M.-C. Tsai, H.-M. Chen, Z. Yang, H. Tan, C.-H. Chen, M.-H. Yeh, A.G. Tamirat, S.-F. Huang, L.-Y. Chen, J.-F. Lee, Y.-F. Liao, E.H. Sargent, H. Dai, and B.-J. Hwang, J. Mater. Chem. A **5**, 21002 (2017).

[16] R.K. Ahrenkiel, Solid. State. Electron. **35**, 239 (1992).

[17] M.C. Nuss and J. Orenstein, in *Millim. Submillim. Wave Spectrosc. Solids* (Springer Berlin Heidelberg, 1998), pp. 7–50.

[18] K.S. Haber and A.C. Albrecht, J. Phys. Chem. **88**, 6025 (1984).

[19] T.J. Savenije, C.S. Ponseca, L. Kunneman, M. Abdellah, K. Zheng, Y. Tian, Q. Zhu, S.E. Canton, I.G. Scheblykin, T. Pullerits, A. Yartsev, and V. Sundström, J. Phys. Chem. Lett. **5**, 2189 (2014).

[20] E. Edri, S. Kirmayer, A. Henning, S. Mukhopadhyay, K. Gartsman, Y. Rosenwaks, G. Hodes, and D. Cahen, Nano Lett. **14**, 1000 (2014).

[21] D. Ritter, E. Zeldov, and K. Weiser, Appl. Phys. Lett. **49**, 791 (1986).

[22] I. Levine, S. Gupta, T.M. Brenner, D. Azulay, O. Millo, G. Hodes, D. Cahen, and I. Balberg, J. Phys. Chem. Lett. **7**, 5219 (2016).

[23] O.E. Semonin, G.A. Elbaz, D.B. Straus, T.D. Hull, D.W. Paley, A.M. van der Zande, J.C. Hone, I. Kymissis, C.R. Kagan, X. Roy, and J.S. Owen, J. Phys. Chem. Lett. **7**, 3510 (2016).

[24] A. Rose, Phys. Rev. **97**, 1538 (1955).

[25] D. Shi, V. Adinolfi, R. Comin, M. Yuan, E. Alarousu, A. Buin, Y. Chen, S. Hoogland, A. Rothenberger, K. Katsiev, Y. Losovyj, X. Zhang, P.A. Dowben, O.F. Mohammed, E.H. Sargent, and

O.M. Bakr, Science **347**, 519 (2015).

[26] N. Ahn, D.-Y. Son, I.-H. Jang, S.M. Kang, M. Choi, and N.-G. Park, J. Am. Chem. Soc. **137**, 8696 (2015).

[27] R.L. Milot, G.E. Eperon, H.J. Snaith, M.B. Johnston, and L.M. Herz, Adv. Funct. Mater. **25**, 6218 (2015).

[28] R.H. Bube, *Photoelectronic Properties of Semiconductors* (Cambridge University Press, 1992).

[29] I. Levine, S. Gupta, T.M. Brenner, D. Azulay, O. Millo, G. Hodes, D. Cahen, and I. Balberg, J. Phys. Chem. Lett. **7**, (2016).

[30] A. Rose, *Concepts in Photoconductivity and Allied Problems* (Interscience, New York, 1963).

[31] H.T. Yi, P. Irkhin, P.P. Joshi, Y.N. Gartstein, X. Zhu, and V. Podzorov, (2018).

[32] N. Kedem, T.M. Brenner, M. Kulbak, N. Schaefer, S. Levcenko, I. Levine, D. Abou-Ras, G. Hodes, and D. Cahen, J. Phys. Chem. Lett. **6**, 2469 (2015).

[33] M.B. Johnston and L.M. Herz, Acc. Chem. Res. **49**, 146 (2016).

[34] L.M. Herz, Annu. Rev. Phys. Chem. **67**, 65 (2016).

[35] R.J. Stoddard, F.T. Eickemeyer, J.K. Katahara, and H.W. Hillhouse, J. Phys. Chem. Lett. **8**, 3289 (2017).

[36] D. Song, P. Cui, T. Wang, D. Wei, M. Li, F. Cao, X. Yue, P. Fu, Y. Li, Y. He, B. Jiang, and M. Trevor, J. Phys. Chem. C **119**, 22812 (2015).

[37] R. Brenes, D. Guo, A. Osherov, N.K. Noel, C. Eames, E.M. Hutter, S.K. Pathak, F. Niroui, R.H. Friend, M.S. Islam, H.J. Snaith, V. Bulović, T.J. Savenije, and S.D. Stranks, Joule **1**, 155 (2017).

[38] K. Sveinbjörnsson, K. Aitola, X. Zhang, M. Pazoki, A. Hagfeldt, G. Boschloo, and E.M.J. Johansson, J. Phys. Chem. Lett. **6**, 4259 (2015).

[39] K. Domanski, W. Tress, T. Moehl, M. Saliba, M.K. Nazeeruddin, M. Grätzel, and M. Gr??tzel,

Adv. Funct. Mater. **25**, 6936 (2015).

[40] C. Momblona, L. Gil-Escrig, E. Bandiello, E.M. Hutter, M. Sessolo, K. Lederer, J. Blochwitz-Nimoth, and H.J. Bolink, Energy Environ. Sci. **9**, 3456 (2016).

[41] Q. Han, Y. Bai, J. Liu, K. Du, T. Li, D. Ji, Y. Zhou, C. Cao, D. Shin, J. Ding, A.D. Franklin, J.T. Glass, J. Hu, M.J. Therien, J. Liu, and D.B. Mitzi, Energy Environ. Sci. **10**, 2365 (2017).

[42] S.S. Lim, W.K. Chong, A. Solanki, H.A. Dewi, S. Mhaisalkar, N. Mathews, and T.C. Sum, Phys. Chem. Chem. Phys. **18**, 27119 (2016).

[43] U. Kwon, M.M. Hasan, W. Yin, D. Kim, N.Y. Ha, S. Lee, T.K. Ahn, H.J. Park, and H.J. Park, Sci. Rep. **6**, 35994 (2016).

[44] C. Li, Q. Guo, W. Qiao, Q. Chen, S. Ma, X. Pan, F. Wang, J. Yao, C. Zhang, M. Xiao, S. Dai, and Z. Tan, Org. Electron. Physics, Mater. Appl. **33**, 194 (2016).

[45] Z. Xiao, Q. Dong, C. Bi, Y. Shao, Y. Yuan, and J. Huang, Adv. Mater. **26**, 6503 (2014).

[46] W. Nie, H. Tsai, R. Asadpour, J.-C. Blancon, A.J. Neukirch, G. Gupta, J.J. Crochet, M. Chhowalla, S. Tretiak, M.A. Alam, H.-L. Wang, and A.D. Mohite, Science **347**, 522 (2015).

[47] D.W. deQuilettes, S.M. Vorpahl, S.D. Stranks, H. Nagaoka, G.E. Eperon, M.E. Ziffer, H.J. Snaith, and D.S. Ginger, Science **348**, 683 (2015).